\documentclass[fleqn,usenatbib]{mnras}

\usepackage{times}

\usepackage[T1]{fontenc}
\usepackage{xcolor}
\usepackage{ae,aecompl}
\usepackage{graphicx}
\usepackage{amsmath}
\usepackage{amssymb}

\title[Shocks in SPH]{Geometrical on-the-fly shock detection in SPH}
\author[Beck, Dolag \& Donnert]{A. M. Beck$^{1}$\thanks{E-mail: abeck@usm.lmu.de}, K. Dolag$^{1,2}$ and J.M.F. Donnert$^{3}$\\
  $^{1}$University Observatory Munich, Scheinerstr. 1, D-81679 Munich, Germany\\
  $^{2}$Max Planck Institute for Astrophysics, Karl-Schwarzschild-Str. 1, D-85741 Garching, Germany\\
  $^{3}$INAF Instituto di Radioastronomia, via P. Gobetti 101, I-40129 Bologna, Italy}

\begin{document}

\label{firstpage}
\pagerange{\pageref{firstpage}--\pageref{lastpage}}

\maketitle


\begin{abstract}
We present an on-the-fly geometrical approach for shock detection and Mach number calculation in simulations employing smoothed particle hydrodynamics (SPH).
We utilize pressure gradients to select shock candidates and define up- and downstream positions.
We obtain hydrodynamical states in the up- and downstream regimes with a series of normal and inverted kernel weightings parallel and perpendicular to the shock normals.
Our on-the-fly geometrical Mach detector incorporates well within the SPH formalism and has low computational cost.

We implement our Mach detector into the simulation code GADGET and alongside many SPH improvements.
We test our shock finder in a sequence of shock-tube tests with successively increasing Mach numbers exceeding by far the typical values inside galaxy clusters.
For all shocks, we resolve the shocks well and the correct Mach numbers are assigned.
An application to a strong magnetized shock-tube gives stable results in full magnetohydrodynamic set-ups.
We simulate a merger of two idealized galaxy clusters and study the shock front.
Shock structures within the merging clusters as well as the cluster shock are well-captured by our algorithm and assigned correct Mach numbers.
\end{abstract}


\begin{keywords}
hydrodynamics -- shock waves -- methods: numerical
\end{keywords}


\section{Introduction}
Shock waves are a common phenomena during the formation and evolution of galaxy clusters.
In the clusters, the intracluster medium (ICM) holds most of its thermal energy in the form of a hot under-dense plasma and shock heating is an important contributor \citep{schindler93,quilis98}.
Shocks within the ICM have been observed with typical Mach numbers of $\approx 1.5 -3$ \citep{markevitch07}.
These shocks are sites of the acceleration of cosmic rays (CR).
The acceleration occurs when charged relativistic particles undergo magnetohydrodynamic fluctuations in the different layers of the shock.

The theory of diffuse shock acceleration has provided a stable framework to describe the acceleration of CR \citep[see e.g.][]{bell78,blandford78,drury81,kang09,caprioli10}.
Therein, the efficiency of particle energy gains contains the Mach number and the magnetic field topology.
The Mach number is the most important ingredient of the efficiency function describing the transfer of kinetic energy into CR acceleration.
In the shocks, CR protons as well as CR electrons are accelerated.
The protons can contribute to the gas pressure and be dynamical important to the evolution of structures.
The electrons, however, cool via synchrotron radiation and can be used as a observational diagnostic tool.
Large-scale shocks within the ICM are also connected to the formation of radio relics in clusters: off-centre giant arcs observed as polarized radio emission \citep[e.g.][]{giov00,vanweeren10,feretti12,bonafede12,brunetti14}.
Therein, shocks accelerate relativistic electrons, which subsequently interact with the magnetic field and emit synchrotron radiation \citep{roettgering97,ensslin98,sarazin99,blasi01,vazza14}.

The field of simulations of large-scale shocks has been pioneered by \cite{miniati00}, who employed Eulerian methods and shock-detecting schemes based on temperature jumps.
Later works adopted more accurate schemes, focusing on the magnitude and distribution of energy dissipation in large-scale environments \citep{ryu03,skillman08,vazza09,vazza11b,schaal15}.
Similar works were performed by \cite{pfrommer06} employing smoothed particle-hydrodynamics (SPH) simulations, with an on-the-fly shock finder based on the change of the entropy with time.
A different method was devised by \cite{hoeft08}, which estimated the Mach number from the entropy ratio caused by the shocks.
However, their formalism is available only in post-processing and computationally expensive.
First implementations of CR processes utilizing the Mach number have been included by \cite{miniati01,ensslin07,pfrommer07,vazza12,kang13,vazza15} within these simulations.

In general, shock-finders in SPH are plagued by noise found within the particle distribution.
This leads to inaccurate Mach numbers and significantly affects the accuracy of the algorithms.
We extend the description of \cite{hoeft08} and present an advanced shock capturing algorithm for SPH to estimate Mach numbers on-the-fly with negligible computational effort.
We show an application to idealized mergers of galaxy clusters, however our implementation is not limited to these scenarios.

This paper is organized as follows.
In Section 2 we present our novel on-the-fly shock detection algorithm.
We perform a sequence of hydrodynamic shock tube tests, a magnetized strong shock and an application to merging galaxy clusters in Section 3 before we summarize in Section 4.


\section{Shock detection method}\label{sec:method}

\begin{figure}
\begin{center}
  \includegraphics[width=0.475\textwidth]{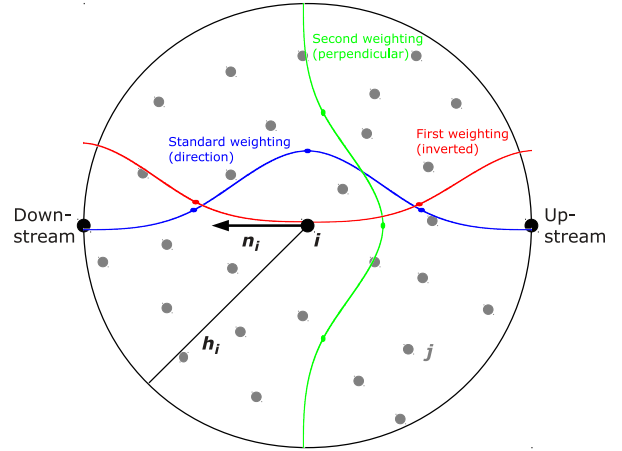}
  \caption{Illustration of the shock detection method.
Considering a particle $i$ with a smoothing length $h_{i}$ and neighbors $j$ we estimate the normal vector $\bmath{n}_{i}$ of a shock candidate with a standard weighting.
Subsequently, we approximate the hydrodynamical states in the up- and downstream regimes by using a product of two weighting.
Firstly, to receive the main contributions only from particles within the up- and downstream regimes we employ an inverted weighting.
Secondly, we account for the distance from the shock normal and employ a perpendicular weighting.}
  \label{fig:shocks}
\end{center}
\end{figure}

We perform the shock detection for each active SPH particle in every time-step as follows (see also Fig. \ref{fig:shocks}).
At first, we estimate the direction of a shock candidate normal vector $\bmath{n}$ and define the corresponding up- and downstream positions $\bmath{x}^{d,u}$.
Next, we approximate the density $\rho^{d,u}$, pressure $P^{d,u}$, velocity $\bmath{v}^{d,u}$ and sound speed $c^{d,u}$ within the up- and downstream regimes.
Subsequently, we calculate the shock properties such as compression ratio $r$, shock speed $v^{sh}$ and Mach number $M$.
At last, we filter for noise and false detections.

\subsection{Shock direction and up- and downstream positions}

At first, we calculate the shock candidate normal vectors.
We perform this calculation alongside the usual SPH density estimation and thus require only little extra computational cost.
Shocks are well characterized by their jumps in the hydrodynamical quantities, such as velocity, pressure, entropy, internal energy or density.
In a proper SPH scheme \citep{beck15} shocks are resolved on the scale of the smoothing length.
For the estimation of the shock direction we need to rely on a well resolved  high signal to noise gradient estimator.
\cite{hoeft08} suggest to define the shock directions by the entropy gradients, however, the entropy shows slight oscillations in our tests.
Thus, we choose the pressure gradients (Hoeft, priv. comm.) as a more reliable definition of potential shock directions.
Therefore, the shock candidate normal vectors $\bmath{n}$ for a particle $i$ pointing into the downstream regimes are given by $\bmath{n}_{i}=-\bmath{\nabla}P_{i}/\left|\bmath{\nabla}P_{i}\right|$.
We calculate the pressure gradients with the SPH estimator \citep[see][]{price12} taking into account the constant error term:
\begin{equation}\bmath{\nabla}P_{i}=\frac{1}{\rho_{i}}\sum_{j}m_{j}\left(P_{j}-P_{i}\right)\bmath{\nabla}_{i}W\left(U_{ij}\right).\end{equation}
For the kernel functions $W$ we use the Wendland $C^{4}$ functions \citep{dehnen12} with 200 neighbors in three dimensions.
We assume the candidate shocks to be spanning across the entire SPH kernel, i.e. the up- and downstream regimes are on the edges of the kernel.
Therefore, the distance of the up- and downstream regimes from the kernel centre is estimated by the smoothing length $\bmath{x}^{d,u}_{i}=\bmath{x}_{i}\pm{}h_{i}\bmath{n}_{i}$.
We can now establish a particle selection criterion: particles within the upstream regime can be selected by $\bmath{n}_{i}\cdot(\bmath{x}_{i}-\bmath{x}_{j})<0$ and particles within the downstream can be selected by $\bmath{n}_{i}\cdot(\bmath{x}_{i}-\bmath{x}_{j})>0$.
This criterion allows us to distinguish the shock regimes and particle contributions in the following summation process.

\subsection{Physical quantities in the up- and downstream regimes}

We estimate the hydrodynamical states within the up- and downstream regimes.
In contrast to \cite{hoeft08}, who require the execution of two extra computational expensive SPH summation loops, we perform an integrated on-the-fly estimation described as follows.
We perform this calculation alongside the usual SPH hydrodynamical force estimation.
Each particle's contribution to the up- and downstream states is calculated with a nested series of weighting distances $U$ (see also Fig. \ref{fig:shocks}).
The standard kernel extension normalized distance scale ($U_{ij}$) between particles $i$ and $j$ is given by
\begin{equation}U_{ij}=\frac{\left|\bmath{x}_{i}-\bmath{x}_{j}\right|}{h_{i}},\end{equation}
which simply corresponds to the inter-particle distance.
We also calculate the distance of the neighbour to a line connecting the estimated positions of the up- and downstream points, which gives a onto the shock normal projected distance from these points towards the central particle by
\begin{equation}U_{ij}^{F}=\left|1-\frac{\left|\bmath{n}_{i}\cdot\left(\bmath{x}_{i}-\bmath{x}_{j}\right)\right|}{h_{i}}\right|,\end{equation}
where we only allow for contributions from within the kernel (i.e. $U_{ij}<1$).
Finally, we also use the perpendicular distance of the particle from the shock normal given by
\begin{equation}U_{ij}^{S}=\frac{\sqrt{\left|\bmath{x}_{i}-\bmath{x}_{j}\right|^{2}-\left|\bmath{n}_{i}\cdot\left(\bmath{x}_{i}-\bmath{x}_{j}\right)\right|^{2}}}{h_{i}}.\end{equation}
The superscripts $F$ and $S$ refer to the namings of the first and second weighting as illustrated in Fig. \ref{fig:shocks}.
For each hydrodynamical quantity we calculate two sums (one for the upstream regime and one for the downstream regime) alongside the hydrodynamical force.
We decide individually to which sum a particle will contribute by employing the selection criteria shown in the previous section.
Particles contributing to the upstream regime are selected by $\bmath{n}_{i}\cdot(\bmath{x}_{i}-\bmath{x}_{j})<0$ and particles contributing to the downstream regime are selected by $\bmath{n}_{i}\cdot(\bmath{x}_{i}-\bmath{x}_{j})>0$.
Thus, we basically split the kernel into two regimes.
The total weight is given by
\begin{equation}w_{ij}=m_{j}^{2}W\left(U^{F}_{ij}\right)W\left(U^{S}_{ij}\right),\end{equation}
where the $W(U)$ terms reflect the weight of the particle because of its distance from the up- or downstream position ($U^{F}$) and of its distance from the shock normal ($U^{S}$).
At last, we normalize the sums and obtain as estimator:
\begin{equation}Z_{i}^{d,u}=\sum_{j^{d,u}}{w_{ij}Z_{j}}\;\;/\;\;\sum_{j^{d,u}}{w_{ij}}\label{signal},\end{equation}
where $Z$ represents a physical quantity and is to replaced with velocity $\bmath{v}$, pressure $P$, density $\rho$ or sound speed $c_{s}$.
The total computational cost of our shock detection method is negligible.

\subsection{Calculation of shock properties}\label{form:mach}

We now calculate the hydrodynamical states at the up- and downstream positions.
At first, we obtain the compression ratio $r$ by a simple division of the corresponding densities $r_{i}=\rho^{d}_{i}/\rho^{u}_{i}$
and the velocity divergence $\Delta{}v$ by projecting the corresponding velocities onto the shock normal
\begin{equation}\Delta{}v_{i}=v^{d}_{i}-v^{u}_{i}=\bmath{n}_{i}\cdot\bmath{v}_{i}^{d}-\bmath{n}_{i}\cdot\bmath{v}_{i}^{u}.\end{equation}
The shock speed $v^{sh}$ then follows 
\begin{equation}v^{sh}_{i}=\frac{\Delta{}v_{i}}{\left|1-\frac{1}{r_{i}}\right|}.\end{equation}
Finally, we calculate the Mach number to $M_{i}=v^{sh}_{i}/c^{u}_{i}$.

\subsection{Filter criteria for noise and false detections}

At last, we filter for particle noise and false detections, caused by noise within the SPH kernel, contributing to the estimation of the shock normal vectors and up- and downstream quantities.
Therefore, we require the shock jumps to exceed tolerance levels of $P^{d}_{i}>\left(1+\varepsilon\right)P^{u}_{i}$ and $\rho^{d}_{i}>\left(1+\varepsilon\right)\rho^{u}_{i}$.
We also require a minimum velocity divergence in the kernel of $\Delta{}v_{i}>0.5\left(1+\varepsilon\right)(v^{d}_{i}+v^{u}_{i})$.
After extensive testing, we settled with an error tolerance criterion of $\varepsilon=0.05$.
For a correct calculation of the velocity divergence and gradient estimators we rely on the advanced SPH scheme of \cite{beck15}.
After a high-order matrix gradient calculation of velocity divergence and curl they employ an improved \cite{balsara95} limiter.
We require this particle-based viscosity limiter $f_i=|\bmath{\nabla}\cdot\bmath{v}|_{i} / (|\bmath{\nabla}\cdot\bmath{v}|_{i} + |\bmath{\nabla}\times\bmath{v}|_{i})$ to be greater than 0.9 and serve as an additional trigger to suppress false detections within the complex network of turbulent and shearing motions.

These values are well suited for the use with an advanced SPH scheme such as the one of of \cite{beck15}, but different variants of SPH might require recalibration efforts.
SPH particles, which fail the filtering process are assigned a Mach number of $M=0$, a compression ratio of $r=1$ and a shock speed of $v^{sh}=0$.
This simple form of filtering is sufficient to suppress fluctuations in the shock detection caused by small-scale noise or shearing motions within the kernel.
As suggested by \cite{hoeft08}, one can also calculate the hydrodynamical states at the four regimes perpendicular to the shock normals, whose orthonormal directions can be constructed with a Gram-Schmidt process.
Afterwards, the velocity divergences perpendicular to the shock can be estimated.
A shock is then selected to be a true shock, if the velocity divergence along the shock larger than the velocity divergences perpendicular to it.
We explored this criterion, but found no improvement over our filtering criteria, besides a significant increase in the amount of computation.


\section{Numerical method and tests}

We implement our Mach detector into the latest version of the GADGET code \citep{springel05}.
For an overview about SPH see \cite{price12}.
We use several state-of-the-art improvements for SPH, which improve its performance in capturing shocks, provide treatments for particle clumping and interpenetration, calculate gradients with higher accuracy and promote mixing between different gas phases \citep[see][for details]{beck15}.

\subsection{Hydrodynamic shock-tube tests}

We perform a sequence of seven standard shock-tube tests \citep{sod78} with Mach numbers ranging from 1.5 to 60.
These tests validate our Mach detector and show its capability to correctly identify well-resolved shock fronts and assign the corresponding Mach numbers.
We consider an ideal gas with an adiabatic index of $\gamma=5/3$, initially at rest.
All our tests are performed in a fully three-dimensional set-up within a periodic box $[x,y,z]$ of dimensions $[140,1,1]$.
We set-up the initial conditions using a relaxed glass-like distribution of $630.000$ particles.
The box is filled with gas of density $\rho_{L}=1$ and of pressure $P_{L}=66.667$ on the left-half side ($x<70$) and with gas of density $\rho_{R}=0.125$ on the right-half side ($x>70$).
We choose the pressure on the right-half side $P_{R}$ such that the resulting solutions yield the Mach numbers $M=\{1.5,\;3.0,\;6.0,\;10.0,\;30.0,\;60.0\}$.

\subsubsection{Series of shock tube tests}

\begin{figure*}
\begin{center}
  \includegraphics[width=0.95\textwidth]{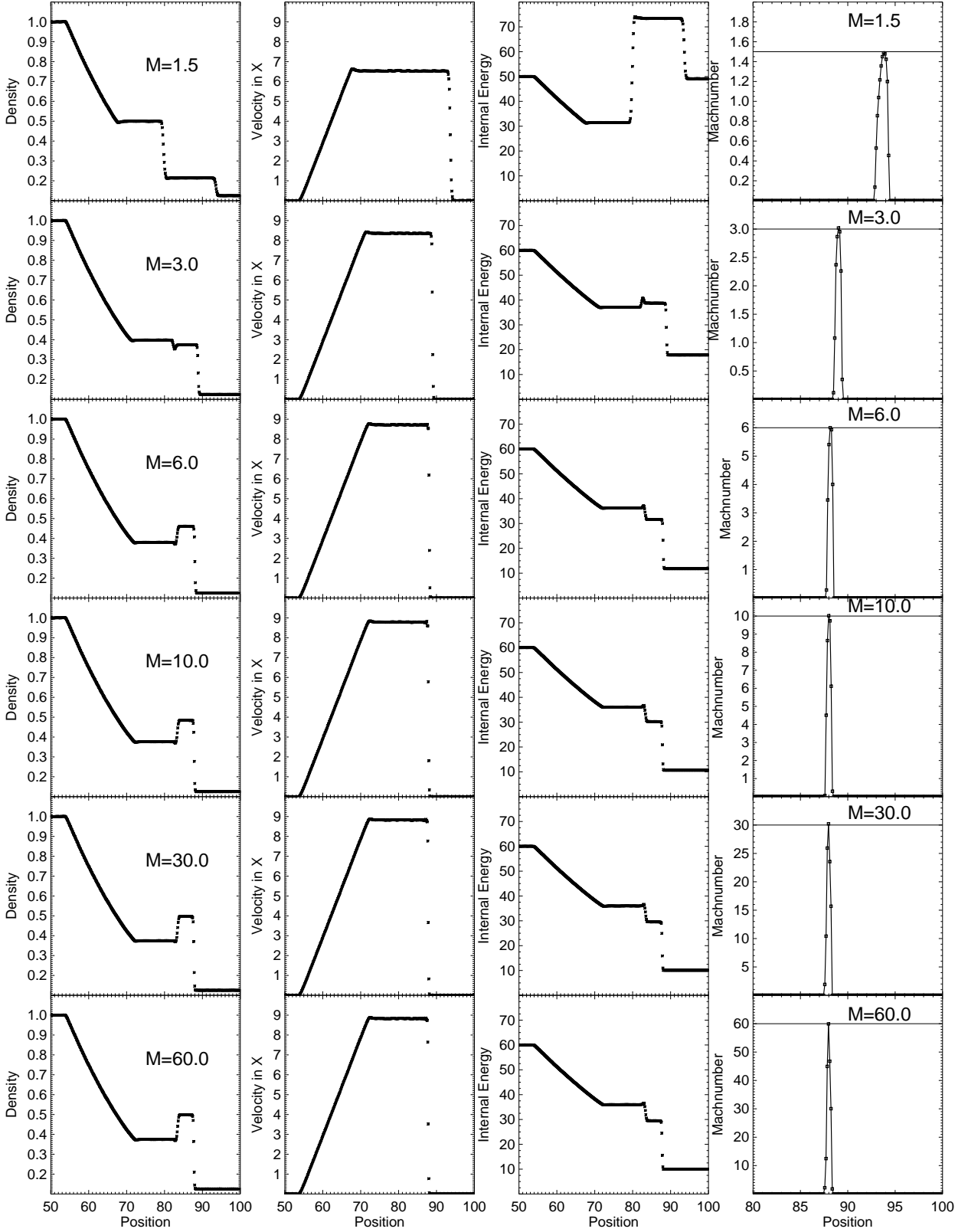}
  \caption{Shock-tube tests with Mach numbers ranging from 1.5 to 60 in a fully three-dimensional set-up.
We show the particle data averaged within bins of half the mean smoothing length.
Our hydrodynamical scheme captures the shocks accurately and the shock detector assigns the correct Mach numbers.}
  \label{fig:machs}
\end{center}
\end{figure*}

Fig. \ref{fig:machs} shows the tests at time $t=1.5$ with well-developed shock fronts.
We volume-average the particle data within bins of half the mean smoothing length and the Mach numbers varies from $1.5$ to $60$ between the top and the bottom panel.
From the left to the right we show gas density $\rho$, $x$-velocity $v_{x}$, internal energy $u$ and Mach number $M$.
Our code is clearly able to capture even the strongest shocks and accurately resolves the corresponding states in density, velocity and internal energy.
However, as intrinsic to the SPH method in the presence of any form of viscosity, the shock fronts are broadened on order of the smoothing length.
The Mach numbers are assigned the correct values and they peak at the locations of the steepest gradients within the shock fronts.
Towards the edges of the shock fronts, the Mach numbers decline and outside the shock fronts, also the non-shocked particles carry the correct Mach number, zero in this case.

Importantly, for our Mach detector to produce accurate results, the underlying SPH method must be able to correctly capture shocks and resolve the corresponding hydrodynamical states.
In particular, the width of the shock is fully determined by the hydrodynamical scheme, where the need for advanced implementations of artificial viscosity arises.
We rely on the improved SPH scheme developed by \cite{beck15}, which uses a high-order gradient matrix scheme to accurately resolve shocks and suppress viscosity in shearing and turbulent motions.

\subsubsection{Zoom-in on the Mach number}

\begin{figure}
\begin{center}
  \includegraphics[width=0.475\textwidth]{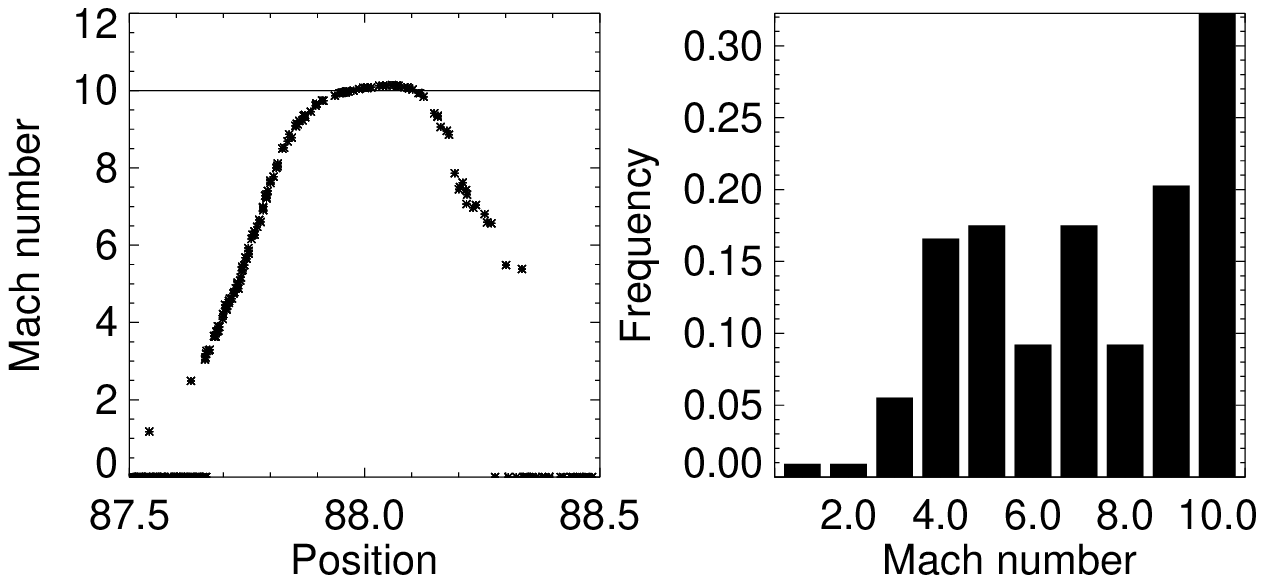}
  \caption{Zoom-in on the Mach number of the M=10 shock tube.
The left panel shows the Mach numbers assigned to individual SPH particles (black dots) compared to the analytical value (horizontal line).
The right panel shows the distribution of non-zero Mach numbers across the shock front.
At the central part, the shock is well captured and large fraction of sufficiently high Mach numbers is computed.
The shock shows the usual SPH broadening, which causes an occurrence of lower numbers.}
  \label{fig:mzoom}
\end{center}
\end{figure}

Fig. \ref{fig:mzoom} shows a zoom-in on the Mach number of the M=10 shock tube.
The left panel shows the raw data, i.e. the Mach numbers assigned to individual SPH particles within the shock front.
The horizontal black line indicates the expected value of M=10.
We find that the central part of the shock is well-captured by our shock detector and all particles carry high Mach values.
Additionally,  however, towards the edges the usual SPH broadening occurs, which leads to a decline of the numbers towards the edges of the shock.
Outside the shock front, particles are assigned values of zero, with a small transition region at the shock outskirts.
The right panel shows the Frequency of non-zero Mach number of particles within the shock.
A large fraction of particles carry high Mach numbers, which corresponds to the central region of the shock.
The fraction of particles with lower values corresponds to the particles sitting at the shock outskirts.

\subsection{Magnetized shock-tube RJ 1A}

We show the magnetized strong shock-tube test '1A' of \cite{ryu95}.
In this test, we enable the full magnetohydrodynamic version of our improved GADGET, which is well described and presented in \cite{dolag09} and \cite{stasyszyn13}.
We consider an ideal gas with an adiabatic index of $\gamma=5/3$, initially at rest.
We perform this test in a fully three-dimensional set-up within a periodic box $[x,y,z]$ of dimensions $[280,1,1]$.
We set-up the initial conditions using a relaxed glass-like distribution of $280.000$ particles.
The box is filled with gas of density $\rho_{L}=1$, pressure $P_{L}=1$, velocity $v_{L}=[10,0,0]$ and magnetic field $B_{L}=[5/(4\pi)^{1/2},5/(4\pi)^{1/2},0]$ on the left-half side ($x<140$) and with gas of density $\rho_{L}=1$, pressure $P_{L}=32.66$, velocity $v_{L}=[-10,0,0]$ and magnetic field $B_{L}=[5/(4\pi)^{1/2},5/(4\pi)^{1/2},0]$  on the right-half side ($x>140$).

\subsubsection{Stable modern SPMHD}

\begin{figure}
\begin{center}
  \includegraphics[width=0.475\textwidth]{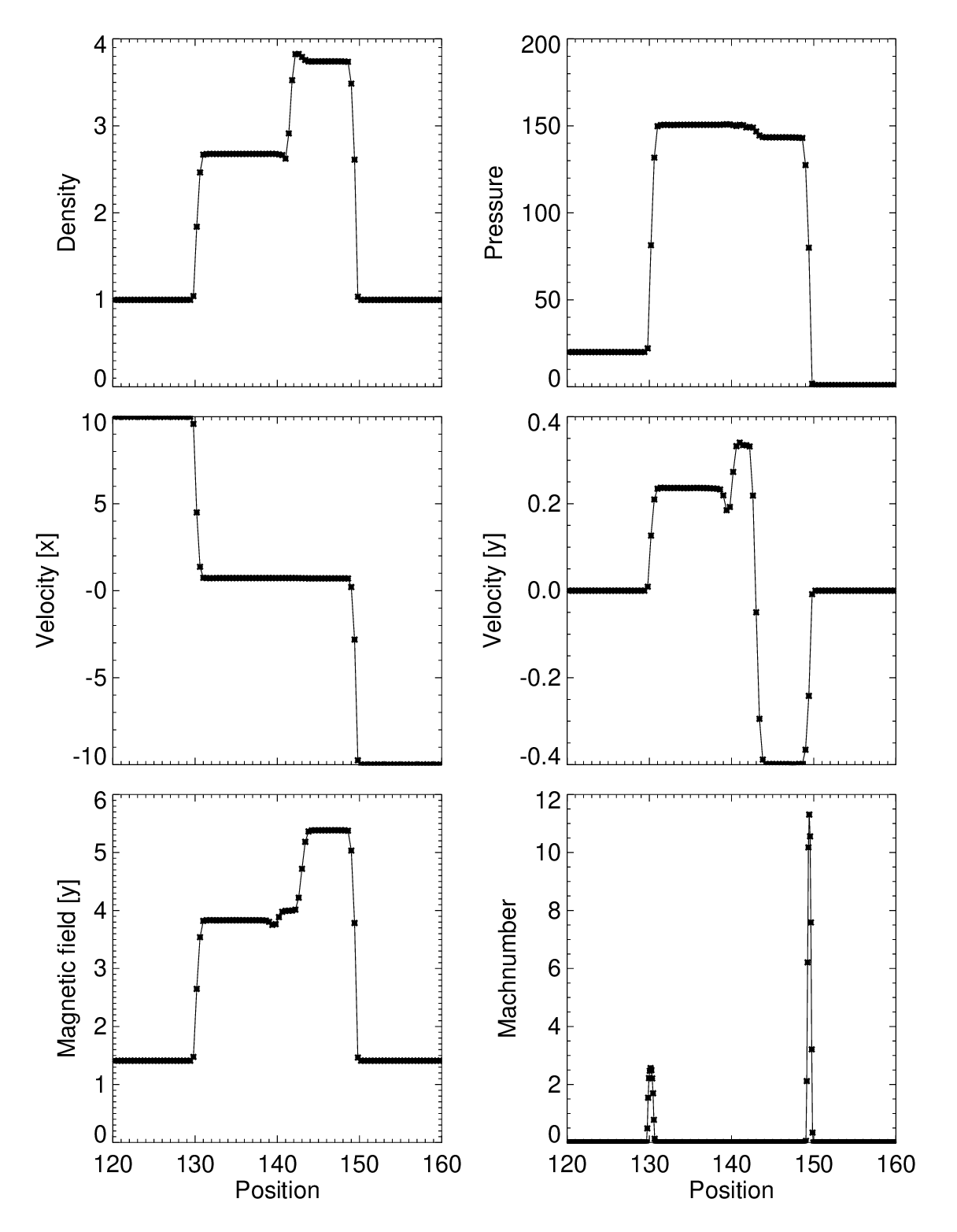}
  \caption{Strong magnetized shock-tube '1A' of Ryu \& Jones 1995.
Our Mach detector performs well in this fully magnetohydrodynamic set-up.
In this test the advantages of an improved formulation of SPMHD are visible.}
  \label{fig:mags}
\end{center}
\end{figure}

Fig. \ref{fig:mags} shows the test at time $t=2.0$ with a well-developed set-up of a fast shock, a slow rarefaction, a contact discontinuity, a slow shock and a fast shock.
This test is usually challenging for magnetohydrodynamic simulations codes, but thanks to our advanced version of GADGET, we obtain a reasonable solution.
Regarding the shocks, it is dominated by hydrodynamical and thermal properties of the gas and the imprint of the magnetic field is only secondary.
Our Mach calculator detects both fast shocks present in the test and associates the corresponding Mach numbers.
The left shock is assigned a Mach number of roughly 2.75 and the right shock of roughly 11.5, again indicating the strong shock behaviour of this test problem.
We note, that we have calculated the sonic Mach number based on the sound speed alone.
This takes into account only the thermal properties of the gas, but is by no means limited towards Alfvenic properties.

\subsubsection{Failure of standard SPMHD}

\begin{figure}
\begin{center} 
  \includegraphics[width=0.475\textwidth]{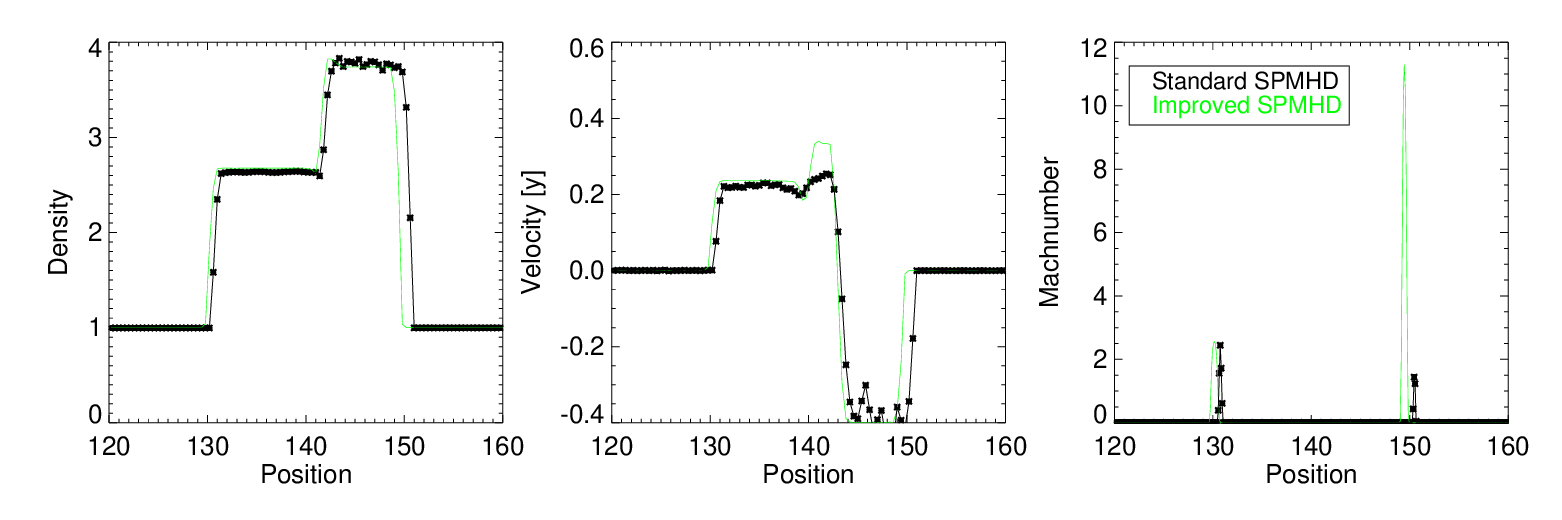}
  \caption{Strong magnetized shock-tube '1A' of Ryu \& Jones 1995 performed with standard SPMHD (black lines) and compared to modern SPMHD (green lines).
We note several difficulties in the computation of the shock such as an off-set of the shock fronts and high levels of noise in density and velocity.
This leads to a failure of the shock finder to properly detect the shock and assign reasonable Mach numbers.}
  \label{fig:magsold}
\end{center}
\end{figure}

Fig. \ref{fig:magsold} shows the same test also at time $t=2.0$, but now computed with a standard formulation of SPMHD.
In this run, we have resorted back to a low-order cubic spline kernel with 64 neighbours and disabled the time-step limiter as well as a higher-order computation of the velocity gradients.
Furthermore, we switched off artificial conduction and only use a primitive version of artificial viscosity.
Basically, all the improvements described in \cite{price12}, \cite{stasyszyn13} or \cite{beck15} have been turned off.
Consequently, the shock is only poorly resolved.
We find an off-set in the shock front as well as high levels of noise in the density estimation and velocity components.
Therefore, the quantities entering the computation of shock normal and Mach numbers also contain inaccurate numbers or high levels of noise.
The Mach detector can only provide a small glimpses and inaccurate numbers.
This test strengthens our claim that for a correct identification of a shock and subsequent computation of its properties, a well-behaved hydro/MHD code is necessary.

\subsection{Two merging galaxy clusters}

We consider the idealized merger scenario of two galaxy clusters as presented in \cite{donnert14}.
In this model, the merger progenitors are set so their main observables match the observed cluster correlations at redshift zero.
The two idealized haloes are based on the Hernquist profile for the DM distribution and a $\beta$-profile for the gas distribution.
The total masses of the clusters are $10^{15}M_{\odot}$ and $5\cdot{}10^{14}M_{\odot}$ and the corresponding virial radii are 2.1 and 1.6 Mpc.
The main cluster is sampled beyond the virial radius up to the size of the periodic box (15471 kpc), giving a natural background DM and gas distribution.
The mass of a dark matter particle is $1.0\times{}10^{8}$ $M_{\sun}$ and of a gas particle $2.8\times{}10^{7}$ $M_{\sun}$, which yields a total of 20 million particles with a gravitational softening of 4.1 kpc.
The velocity of the collision-less particles is set by numerically solving the Eddington equation, taking into account the gravitational potential generated by the gas in the cluster \citep[see][for details]{donnert14}.
We separate the individual centre of masses of both clusters by a distance of 3.7 Mpc and the initial approach velocity of the clusters is 1900 km/s.
Our scheme promotes the mixing of gas phases and produces entropy cores in contrast to previous SPH with declining entropy profiles \citep{vazza11a,beck15}.

\begin{figure}
\begin{center}
  \includegraphics[width=0.475\textwidth]{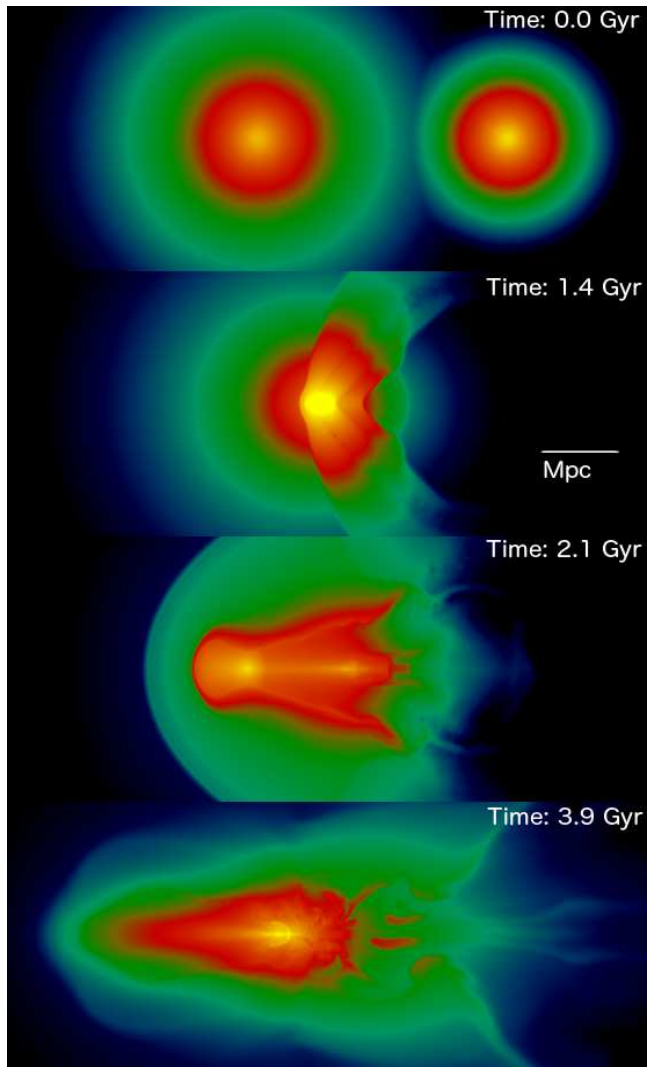}
  \caption{Idealised merger of two galaxy clusters.
We show the gas density in thin slices through the centre of the clusters.
The various stages of the merger event (infall, shock waves, core oscillations and turbulence) are visible.}
  \label{fig:merger}
\end{center}
\end{figure}

Fig. \ref{fig:merger} shows the time evolution of the gas density during the merger process.
In principle, the merger is the smaller cluster penetrating into the larger cluster causing several occurrences of shocks:
An accretion shock during first infall, an outward-traveling front shock after infall and an outward-traveling reverse shock after infall.
Additionally, several weak shocks are present in the most inner regions as both cluster cores oscillate during relaxation.

\begin{figure}
\begin{center}
  \includegraphics[width=0.475\textwidth]{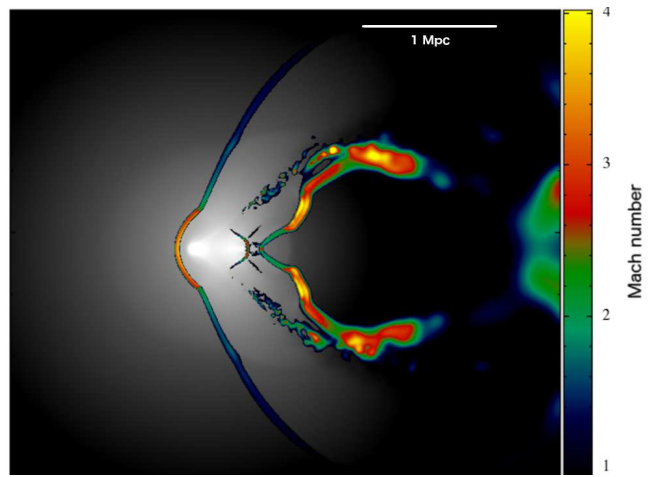}
  \caption{Formation of the shock front at time 1.5 Gyr.
The relic emerging from the merger (left curved shock) as well as several X,M or W-like shocks in the centre and a reverse shock are visible.
We overlay a thin slice of the gas density (grey-scale) with a thin slice of the Mach number (color).}
  \label{fig:small}
\end{center}
\end{figure}

Fig. \ref{fig:small} shows the formation of the shock front at time t = 1.5 Gyr.
This plot illustrates the ability of our Mach detector to not only capture the main shock front, i. e. relic, caused by the cluster merger, but also several small structures.
Furthermore, also the reverse shock can be identified.
A study of the small-scale shock structures and a careful separation between noise and turbulent structures is beyond the scope of this paper.
We leave a detailed analysis and comparison with observations of radiation mechanisms to a future work.

\begin{figure}
\begin{center}
  \includegraphics[width=0.475\textwidth]{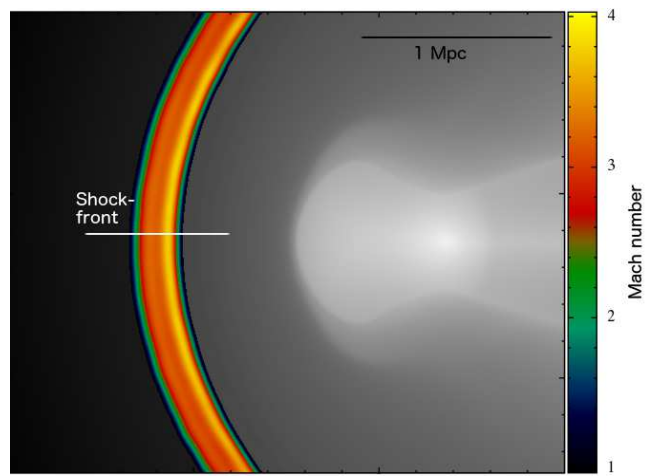}
  \caption{Zoom-in on the cluster merger relic at a time 2.1 Gyr.
We overlay a thin slice of the gas density (grey-scale) with a thin slice of the Mach number (color).
The horizontal white line illustrates our cut through the shock front (see Fig. \ref{fig:mach_zoom}).}
  \label{fig:mach_cut}
\end{center}
\end{figure}

Fig. \ref{fig:mach_cut} shows a zoom-in on the cluster relic shock front at time t = 2.1 Gyr.
We overlay a thin slice of the gas density (grey-scale) with a thin slice of the Mach number (color).
The shock front has the shape of a spherical cap, with radius of roughly three Mpc, a total width of about 300 kpc and a peak Mach number between 3 and 4.
The Mach number increases smoothly towards the peak from the subsonic regimes before and behind the shock.
As the shock front moves further outwards and thereby propagates into a lower density medium, the Mach number increases.
Early on, the Mach number is roughly 3 and later becomes as large as 5.

\begin{figure}
\begin{center}
  \includegraphics[width=0.475\textwidth]{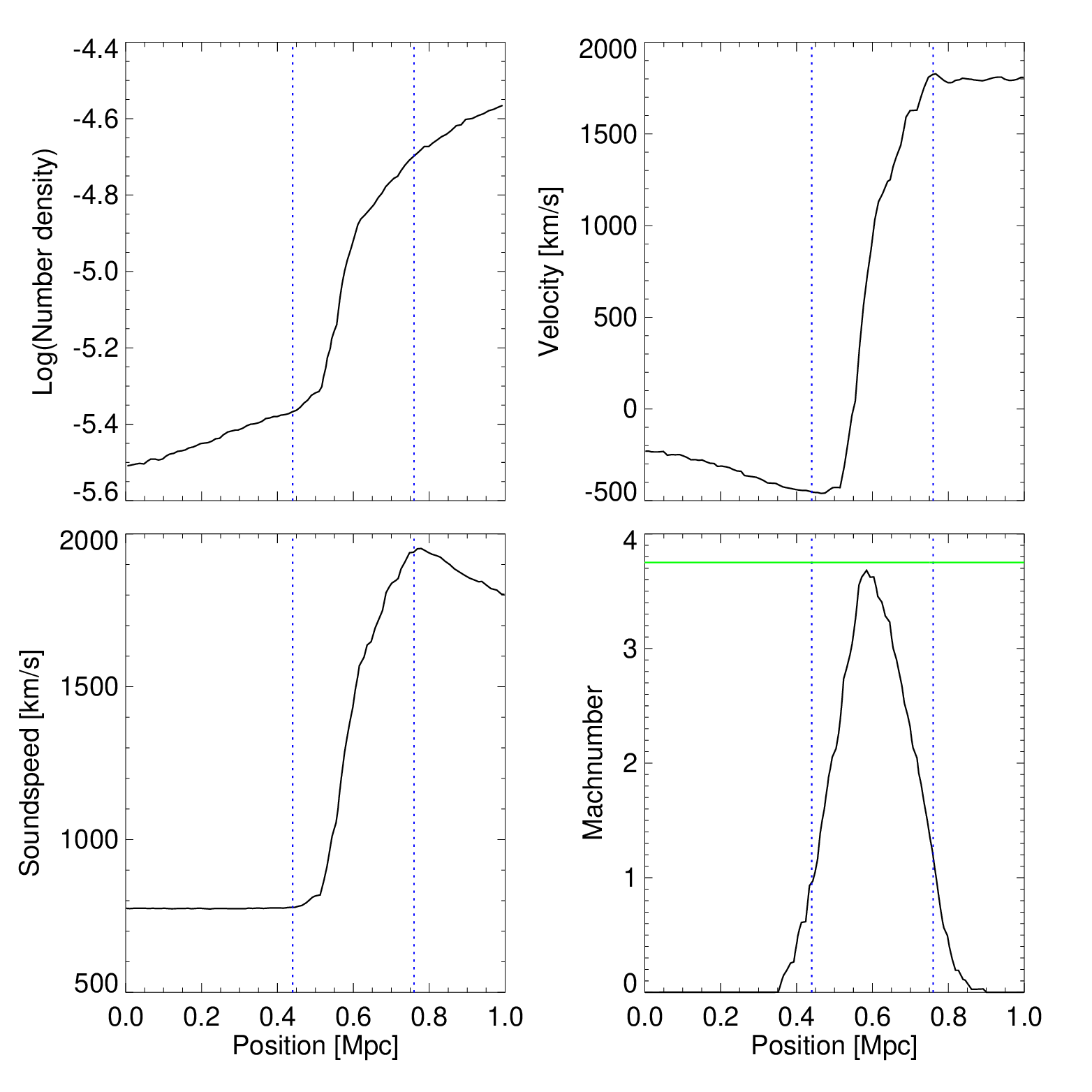}
  \caption{Cut through the shock front at time 2.1 Gyr.
The shock front and the corresponding jumps in the hydrodynamical states are clearly visible.
The Mach number peaks within the shock front at M = 3.75 and besides broadening contains very little noise.
The horizontal green line in the bottom right panel illustrates the manually calculated Mach number (at the positions of the vertical blue dotted lines), which compares well to the peak Mach number extracted from the simulation.
The vertical blue lines do also indicate the width of the SPH kernel ($\approx 320$ kpc) around the peak Mach number.}
  \label{fig:mach_zoom}
\end{center}
\end{figure}

Fig. \ref{fig:mach_zoom} shows the hydrodynamical states in a cut across the shock front.
The bottom right panel shows the on-the-fly during the simulation calculated Mach number across the shock front.
The top left panel, the bottom left panel and the bottom right panel show the gas number density, sound speed and velocity (projected onto the shock normal).
We have calculated these quantities by averaging the particle data within a small segment parallel to the shock front.
From these quantities we calculate the Mach number.
We estimate for the up- and downstream quantities (taken at the positions of the vertical blue lines indicated in the panels):
$c_{s}^{u}=$ 779 km/s, $c_{s}^{d}=$ 1935 km/s, $n^{u}=$ 4.35$\cdot10^{-6}$, $n^{d}=$ 1.95$\cdot10^{-5}$, $v^{u}=$ -458 km/s, $v^{d}=$ 1811 km/s.
The manually calculated Mach number peaks at M=3.75 (horizontal green line in the bottom right panel) and agrees well with the on-the-fly estimation.
The locations, which we have used for the manual calculation are consistent with the width of the SPH kernel around the peak Mach number.
The mean smoothing length is of the order 160 kpc and is given by the employed kernel function, which then yields a kernel width of roughly 320 kpc.
In this run we have used the Wendland $C^{4}$ kernel with 200 neighbors in three dimensions.


\section{Summary}

We presented a geometrical on-the-fly method for shock detection and Mach number calculation in SPH simulations.
The shock detector is based on suggestions of \cite{hoeft08}, but adapted to be used on-the-fly.
We perform a geometrical weighting approach in contrast to previous works in SPH \citep[see e.g.][]{pfrommer06}, who used the rate of change of entropy to characterize shocks.
Furthermore, in our formalism SPH particles are evaluated at all times and we do not require to lock their Mach numbers for certain time intervals.

We applied our Mach detector to a sequence of hydrodynamical shock tubes with Mach numbers ranging from $M=1.5$ to $M=60$.
Our Mach calculator performs robustly and accurately in the weak and strong shock regime.
Furthermore, we analyzed the idealized relic shock caused by the merger of two galaxy clusters with a mass ratio of 2:1.
We find the relic to be roughly of $M=3.75$ with a width of order 300 kpc and radius of three Mpc.

Importantly, for our Mach detector to produce accurate results, the underlying SPH method must be able to correctly capture shocks and resolve the corresponding hydrodynamical states.
In particular, the width of the shock is fully determined by the hydrodynamical scheme, where the need for advanced implementations of artificial viscosity arises.
We rely on the improved SPH scheme developed by \cite{beck15}, which uses a high-order gradient matrix scheme to accurately resolve shocks and suppress viscosity in shearing and turbulent motions.
At last, we are able to properly resolve the full shock dynamics and relics caused by mergers of galaxy clusters.

Our Mach detector can be easily employed within future large-scale simulations and allows an analysis of shock properties in post-processing as well as on-the-fly for calculations of energy injection and particle acceleration.
We have also performed some first fully magnetohydrodynamic shock-tubes and see stable and promising results.
Within this test, in particular, we see the need for a well-behaved hydro/MHD scheme arising.
With the knowledge of the distribution, statistics and time evolution of shocks, the dynamical contribution of CR protons can be estimated and improvements on the estimation of synthetic radio observables are possible.
It can be applied to estimate the synchrotron emission of large Radio Relics caused by galaxy clusters, or of Radio Haloes caused by subsonic turbulent motions.
Additionally, the radio emission of the cosmic web can be approximated and compared with future surveys.


\section*{Acknowledgments}
AMB thanks the Osservatorio Astronomico di Trieste for its hospitality during a long-term stay in 2015.
We thank Franco Vazza, Susana Planelles and Giuseppe Murante for very useful discussions.
We thank Tadziu Hoffmann for his help in handling large sets of data.
We thank Volker Springel for access to the private version of the GADGET code.
Images were created with the SPLASH software \citep{price07}.
AMB and KD are supported by the DFG Research Unit 1254 'Magnetization of interstellar and intergalactic media', the DFG Cluster of Excellence 'Origin and Structure of the Universe' and the DFG TransRegion 33 'The Dark Universe'.
JMFD is supported by the EU FP7 Marie Curie program 'People'.


\bibliographystyle{mnras}
\bibliography{shocks.bib}

\bsp

\label{lastpage}

\end{document}